\documentclass[11pt]{article}
\usepackage[dvips]{graphics}
\usepackage{graphicx}
\title{Classical mechanics as nonlinear quantum mechanics}
\author{Hrvoje Nikoli\'c \\
Theoretical Physics Division, Rudjer Bo\v{s}kovi\'{c} Institute, \\
P.O.B. 180, HR-10002 Zagreb, Croatia \\
{\normalsize e-mail: hrvoje@thphys.irb.hr} \\
\makebox[1in]{} \\
}
\date{\today}
\begin{document}
\maketitle
\begin{abstract}
All measurable predictions of classical mechanics can be reproduced
from a
quantum-like interpretation of a nonlinear Schr\"odinger equation.
The key observation leading to classical physics is the fact that a
wave function that satisfies a linear equation is real and positive,
rather than complex. This has profound implications on the role of
the Bohmian classical-like interpretation of linear quantum mechanics,
as well as on the possibilities to find a consistent interpretation
of arbitrary nonlinear generalizations of quantum mechanics.
\end{abstract}

\maketitle

\section{Introduction}

In physics, linear equations are often approximations to nonlinear equations.
On the other hand, we know that the Schr\"odinger equation is a 
linear equation. This raises an interesting question:
Does it mean that the Schr\"odinger equation could be an approximation 
to a nonlinear equation?

We also know that the interpretation of 
the Schr\"odinger equation strongly rests on linearity.
Therefore, the possibility of a nonlinear generalization
of the Schr\"odinger equation raises further questions, such as:
Does it mean that we have to modify the interpretation of 
quantum mechanics (QM)?
Can nonlinearities teach us something new about the interpretation of QM?

Some results in that direction have already been found.
In \cite{gis,pol,cza} the following result has been obtained:
{\em If (nonlinear) wave function collapses, then EPR correlations
can be used to transmit information instantaneously.}
(For a comparison, it is well known that in ordinary linear QM
the EPR correlations cannot be used to transmit information
instantaneously.)
This suggests that the concept of a true wave-function collapse
is problematic. An alternative is to adopt an interpretation that
does not rest on a true wave-function collapse. An example
of such an interpretation is the many-world interpretation.
However, when the many-world interpretation is applied to
the nonlinear case, it is found that then 
different branches (worlds) may communicate \cite{pol}.
But if they communicate, then it does not seem meaningful
to interpret them as different worlds.

We see that the properties of nonlinear QM seem rather pathological.
This suggests that a radically different interpretation should be 
adopted. 
In particular, interpretations in which the wave function
is an objective-realistic entity are not expected to lead to such  
pathologies, as such interpretions are more similar to
classical nonlinear waves (which, of course, do not lead to
interpretational pathologies).

To resolve the interpretational problems, it would be desirable to have 
at least one example of nonlinear QM which we know how to 
interpret. On the other hand, it is widely believed that no such 
example of nonlinear QM exists. Nevertheless, we show
that such an example exists, i.e., that {\em classical mechanics} (CM)
represents an example of nonlinear QM.  
More precisely, we review our results of \cite{nik} which 
show that classical mechanics can be represented
by a nonlinear Schr\"odinger equation
\begin{equation}\label{schcl}
\left[ \frac{\hat{{\bf p}}^2}{2m} + V- Q \right] \psi =i\hbar\partial_t\psi,
\end{equation}
where
\begin{equation}\label{QQ}
Q\equiv -\frac{\hbar^2}{2m}\frac{\nabla^2 |\psi|}{|\psi|} .
\end{equation}
We see that $Q$ in (\ref{QQ}) explicitly depends on $\psi$, which makes
(\ref{schcl}) nonlinear in $\psi$.
We also note that the classical Schr\"odinger equation (\ref{schcl})
has also been discussed earlier (see, e.g., \cite{ros}), but that before 
\cite{nik} it has not been shown that this equation alone is 
sufficient to reproduce {\em all} measurable predictions of the usual
formulation of CM. 

Another motivation for viewing CM as nonlinear QM is a hope
that it could help in a better understanding of the big conceptual difference
between CM and QM. For example, 
where does this difference comes from?
A frequent answer is that this
difference comes from the fact that CM is deterministic, 
whereas QM is probabilistic.
Nevertheless, there is a deterministic interpretation of QM very 
similar to CM, but with the same probabilistic predictions as standard 
QM -- the Bohmian interpretation \cite{bohm}. 
Then why the Bohmian interpretation is not widely accepted?
A frequent answer is: Because the purely probabilistic interpretation 
is simpler. 
(Or more precisely, because the Bohmian trajectories are unobservable, 
and thus unnecessary, hidden variables.)
If we accept this argument against the Bohmian interpretation,
then the following natural question arises:
Can CM be made simpler by adopting a probabilistic interpretation?
As we shall see in this paper, the answer will turn out to be -- yes.
We shall see that
the probabilistic QM-like interpretation of the classical nonlinear 
Schrodinger equation reproduces all measurable predictions of standard CM.
We shall also see that
classical trajectories existing even without measurements play a role of 
unobservable hidden variables analogous to the Bohmian trajectories in QM.
Thus, 
if it is (un)natural to accept the Bohmian interpretation of QM, then 
it is equally (un)natural to accept classical trajectories of CM.

\section{QM and the Bohmian interpretation}

Consider the standard Schr\"odinger equation
\begin{equation}\label{sch}
\left[ \frac{\hat{{\bf p}}^2}{2m} + V({\bf x},t) \right] \psi({\bf x},t)
=i\hbar\partial_t\psi({\bf x},t),
\end{equation} 
where 
\begin{equation}\label{p}
\hat{{\bf p}}=-i\hbar\nabla .
\end{equation}
We write $\psi$ in the polar form
\begin{equation}\label{psi}
\psi({\bf x},t)=R({\bf x},t)e^{iS({\bf x},t)/\hbar} ,
\end{equation}
where $R$ and $S$ are real functions and
\begin{equation}\label{R>0}
R({\bf x},t)\geq 0.
\end{equation}
The complex Schr\"odinger equation is equivalent to 
a set of 2 real equations. These are
the quantum Hamilton-Jacobi equation
\begin{equation}\label{HJQ}
\frac{(\nabla S)^2}{2m}+V+Q=-\partial_tS,
\end{equation}
and the conservation equation
\begin{equation}\label{cons}
\partial_t\rho + \nabla\left( \rho\frac{\nabla S}{m} \right) =0,
\end{equation}
where
\begin{equation}\label{Q}
Q\equiv -\frac{\hbar^2}{2m}\frac{\nabla^2 R}{R}.
\end{equation}
Similarity with the classical Hamilton-Jacobi equation suggests 
the Bohmian interpretation \cite{bohm}. In this interpretation, 
the particle has a trajectory satisfying
\begin{equation}\label{bohm}
\frac{d {\bf x}}{dt}=\frac{\nabla S}{m} ,
\end{equation}
which is the same as an analogous equation in 
the classical Hamilton-Jacobi theory.
The conservation equation provides that particles in a statistical ensemble 
are always distributed as in QM, with the probability
density $\rho({\bf x}, t)$.
Thus, in the Bohmian interpretation, 
all QM uncertainties emerge from the lack of knowledge of the actual 
initial particle position ${\bf x}(t_0)$.

\section{Classical Schr\"odinger equation}

Now consider a classical statistical ensemble in the configuration space.
The conservation equation is the same as in QM, which can be written
as a linear equation for $R$
\begin{equation}\label{lin}
\left[ \partial_t +\left( \frac{\nabla S}{m} \right) \nabla
+ \left( \frac{\nabla^2 S}{2m} \right) \right] R =0 ,
\end{equation}
where $R\equiv\sqrt{\rho}$.
The classical Hamilton-Jacobi equation reads
\begin{equation}\label{HJ}
\frac{(\nabla S)^2}{2m}+V=-\partial_tS.
\end{equation}
Defining the classical wave function as
\begin{equation}\label{psirepeat}
\psi({\bf x},t)=R({\bf x},t)e^{iS({\bf x},t)/\hbar} ,
\end{equation}
the conservation equation and the Hamilton-Jacobi equation 
together turn out to be equivalent to the nonlinear classical 
Schr\"odinger equation
\begin{equation}\label{schclrepeat}
\left[ \frac{\hat{{\bf p}}^2}{2m} + V- Q \right] \psi =i\hbar\partial_t\psi .
\end{equation}
If, in addition, the wave function is required to be single-valued, 
then CM improves by 
including the Bohr quantization condition \cite{nik}
\begin{equation}
mvr=n\hbar .
\end{equation}

\section{Measurement in nonlinear QM}

We start from the observation that
any function $\psi({\bf x},t)$ 
representing a solution of some (not necessarily linear) equation
can be expanded in terms of some other functions as
\begin{equation}\label{wf}
\psi({\bf x},t)=\sum_a c_a\psi_a({\bf x},t) . 
\end{equation}
The linear case is special by having the property
that the functions $\psi_a$ can be chosen such that
$c_a\psi_a$ and $\psi_a$ are also solutions.
Now, what is a measurement?
Typically, a measurement is a process in which
we obtain knowledge that the actual state is $c_a\psi_a$. 
Thus, to determine the subsequent post-measurement 
properties of the system, it is sufficient to know only 
that component. 
However, in the nonlinear case, it is not a solution, so 
to know that component one actually needs to know
the {\em whole} solution.
Only in the linear case the measured component evolves independently of 
other components.
This explains the {\em effective} collapse in the linear case.

Now consider a measuring apparatus. Let the
eigenstates of a measured hermitian operator be $\psi_a({\bf x},t)$.
A measurement that does not disturb the wave function requires 
entanglement with the measuring apparatus,
such that the total wave function takes a form
\begin{equation}\label{wfy}                              
\Psi({\bf x},{\bf y},t)=\sum_a c_a\psi_a({\bf x},t)\phi_a({\bf y},t),
\end{equation}
where the coefficients $c_a$ are the same as those in (\ref{wf}) 
and $\phi_a({\bf y},t)$ are some orthonormal states of the measuring
apparatus.
However, in the general nonlinear case, such a solution does not exist.
From this, we conclude that
{\em it is much more difficult to measure a quantity in nonlinear QM than in 
linear QM}.

\section{Measurement for the classical Schr\"odinger equation}

The final conclusion of the preceding paragraph raises the following
question:
Does it mean that it is very difficult to measure anything in the 
quantum theory described by the classical Schr\"odinger equation?
Fortunately, the answer is -- no! Instead, as we show below,
the classical Schr\"odinger equation has some specific properties that 
allow measurements that reproduce standard CM.

Although $\psi$ does not satisfy a linear equation, there 
is a quantity that {\em does} satisfy a linear equation.
This quantity is $R({\bf x},t)$.
(Note that for the linear Schr\"odinger equation 
$R$ is not determined by a linear equation 
because $R$ appears also in the quantum Hamilton-Jacobi equation.)
Thus,
{\em for measurement and effective collapse, the relevant ``wave function" 
is $R({\bf x},t)$.}
The positivity of this wave function will turn out to be the source of 
classical properties that emerge from the classical Schr\"odinger equation.

First, in analogy with standard QM, we introduce the notation
\begin{equation} 
R({\bf x})=\langle{\bf x}|R\rangle=\langle R|{\bf x}\rangle ,
\end{equation}
where the last equality is a consequence of reality.
The scalar product is
\begin{equation}
\langle R_1| R_2\rangle \equiv \int d^3x 
\langle R_1|{\bf x}\rangle\langle{\bf x}| R_2\rangle
\equiv\int d^3x \, R_1({\bf x}) R_2({\bf x}) ,
\end{equation}
which is {\em positive} (i.e., real and nonnegative).
The only complete orthogonal basis $\{ |R_i\rangle \}$     
consistent with the positivity requirement
$\langle {\bf x}|R_i\rangle\geq 0$ is the position basis 
$\{ |{\bf x}\rangle \}$.
Thus, {\em the position basis is the preferred basis}.
The most general state consistent with the positivity requirement is
\begin{equation}
|R\rangle = \int d^3x \, c({\bf x}) |{\bf x}\rangle ,
\end{equation}
where $c({\bf x})\geq 0$. 

We see that
no real state $R({\bf x})$ is an eigenstate of the momentum operator.
Consequently, the state cannot collapse to the momentum eigenstate.
This implies that the Heisenberg uncertainty relation 
$\Delta x \Delta p \geq\hbar/2$
cannot be revealed by an experiment, in agreement with the fact that
there is no Heisenberg uncertainty relation in classical mechanics.
Still, it does not mean that momentum cannot be measured. Instead,
momentum can be measured {\em indirectly} by measuring two subsequent 
positions ${\bf x}_1$, ${\bf x}_2$
at times $t_1$, $t_2$, respectively.
The momentum is then defined as
\begin{equation}
{\bf p}=({\bf x}_2-{\bf x}_1)/m(t_2-t_1) .
\end{equation}
This, indeed, is how momentum is measured in classical mechanics.

\section{Emergence of classical statistics}

The origin of all nonclassical (i.e., typically quantum) probabilistic phenomena 
(e.g. destructive interference, EPR correlations, 
violation of Bell inequalities, ...) 
can be traced back to the fact that the scalar product 
$\langle \psi_1| \psi_2\rangle$
between the probability amplitudes does not need to be positive.
Therefore, the positivity implies that there is no such nonclassical 
probabilistic phenomena in CM.

Furthermore, particles can always be distinguished. To see this,
consider a 2-particle state
\begin{equation}
R({\bf x},{\bf y})=R_1({\bf x})R_2({\bf y})+R_2({\bf x})R_1({\bf y}) ,
\end{equation}
where $R_1$ and $R_2$ are {\em orthogonal}.
Consequently, in the probability density $R^2$,
the exchange term vanishes
\begin{equation}
2R_1({\bf x})R_2({\bf x})R_1({\bf y})R_2({\bf y}) =0 .
\end{equation}
Just as in standard QM, this means that two particles can be regarded as 
distinguishable.

Now consider the density matrices.
A pure state
\begin{equation}\label{pures}
|R\rangle=\sum_{i} \sqrt{w_i} |R_i\rangle ,
\end{equation}
where $w_i \geq 0$, 
can be represented by a density matrix
\begin{equation}
\hat{\rho}_{\rm pure}=|R\rangle\langle R| .
\end{equation}
The associated mixed state is a diagonal state
\begin{equation}
\hat{\rho}_{\rm mix}=\sum_{i} w_i |R_i\rangle \langle R_i|  .
\end{equation}
The pure and the mixed state are related as
\begin{equation}
\hat{\rho}_{\rm pure}=\hat{\rho}_{\rm mix} +
\sum_{i\neq j} \sqrt{w_i w_j} 
|R_i\rangle \langle R_j| .
\end{equation}
Is there a measurable difference between pure and mixed states?
Since $\hat{{\bf p}}$ is not measurable, the most general measurable 
operator is $\hat{A}=A({\bf x})$,     
which is diagonal in the preferred basis. Consequently
\begin{equation}
\langle \hat{A}\rangle = {\rm Tr} (\hat{\rho}_{\rm mix} \hat{A})
= {\rm Tr} (\hat{\rho}_{\rm pure} \hat{A})
= \int d^3x\, \rho({\bf x}) A({\bf x}) ,
\end{equation}
where $\rho({\bf x})=R^2({\bf x})$ and 
$R({\bf x})=\sum_i \sqrt{w_i} R_i({\bf x})$.
This means that {\em there is no measurable difference between pure states 
and the associated mixed states}.
The off-diagonal part plays no measurable role. Effectively it does not 
appear, which is a property of classical statistical mechanics. 
(Note a similarity with decoherence in ordinary QM).

\section{Emergence of classical trajectories}

The next question is:
Why particles appear to move along classical trajectories?
A partial answer is provided by the
Ehrenfest theorem valid for the classical Schr\"odinger equation.
One can show that
\begin{equation}\label{Eh1}
\langle {\bf x}\rangle =\int d^3x \, \psi^*({\bf x},t) {\bf x} 
\psi({\bf x},t)=\int d^3x \, \rho({\bf x},t) {\bf x} ,
\end{equation}
\begin{equation}\label{Eh2}
\frac{d\langle {\bf x}\rangle}{dt} =\int d^3x \, \rho \frac{\nabla S}{m}
=\int d^3x \, \psi^* \frac{\hat{{\bf p}}}{m} \psi ,
\end{equation}
\begin{equation}\label{Eh3}
m\frac{d^2\langle {\bf x}\rangle}{dt^2} =\int d^3x \, \rho 
(-\nabla V)=\int d^3x \, \psi^* (-\nabla V)\psi .
\end{equation}
Thus, as in ordinary QM, the Ehrenfest theorem says that
the average position satisfies classical equations 
of motion, but that the actual position may be uncertain.
But why particles appear as pointlike?
As in ordinary QM, a measurement of the position induces a collapse of 
$R({\bf x})$ to an arbitrarily narrow wave packet.
However, in linear QM, we know that the wave packet suffers dispersion, i.e.,   
that narrow wave packets are not stable.
By contrast, the classical nonlinear Schr\"odinger equation contains 
arbitrarily narrow stable soliton solutions. The point-particle 
soliton solution is \cite{nik} 
\begin{equation}\label{xp}
\psi_{\rm sol}({\bf x},t)=\sqrt{ \delta^3({\bf x}-{\bf y}(t)) } \,
e^{iS({\bf x},t)/\hbar} ,
\end{equation}
where
\begin{equation}\label{bohm'}
\frac{d{\bf y}(t)}{dt} =\frac{\nabla S({\bf y}(t),t)}{m} .
\end{equation}
Note that (\ref{bohm'}) does not describe the motion of a 
pointlike particle associated with any solution of the
classical Schr\"odinger equation, but the motion of the 
crest of the wave packet.

The results above can be interpreted as follows:
{\em If a particle is measured to have a definite position at some time, 
then it will remain to have a definite position at later times and 
this position will change with time according to classical equations 
of motion.}
However, {\em if the particle is not measured to have a definite position, 
then one is not allowed to claim that the particle has  a definite position}.
While this interpretation contradicts the usual interpretation of CM, 
it is analogous to the reasoning in the usual interpretation of QM and does 
not contradict any measurable result of CM.

We also note that, at each time, both a position and a momentum
can be associated with such soliton solutions. This allows
to introduce a classical phase space for this nonlinear QM \cite{nik}.

\section{Discussion}

Now we see that 
there are 4 consistent ways to interpret CM and QM:
\begin{enumerate}
\item Traditional: CM is deterministic, QM is probabilistic.
\item Bohmian: Both CM and QM are deterministic.
\item Anti-Bohmian: Both CM and QM are probabilistic.
\item Anti-traditional: CM is probabilistic, QM is deterministic.
\end{enumerate}
The natural question is: How to know
what is the correct way to interpret CM and QM?
The Occam's razor says -- the simplest one!
But what does it mean ``the simplest"?
We have to chose some criterion of simplicity.
With two different criteria we obtain two different answers:
\begin{enumerate}
\item Technical simplicity: no guiding equation for a particle trajectory -- 
implies that both QM and CM are probabilistic.
\item Conceptual simplicity: particle positions exist even if we do not 
measure them -- implies that both QM and CM are deterministic.
\end{enumerate}
This suggests that we should either: 
\begin{enumerate}
\item reject determinism of CM, or
\item accept the Bohmian deterministic interpretation of QM.
\end{enumerate}
The interesting thing is that both possibilities seem rather heretic. 

The next question we study is the following:
Is there a consistent interpretation of nonlinear QM (in general)?
From our results, we can conclude that
if the consistent theory of measurement is the crucial consistency 
requirement, then the answer is -- no! 
This is because we have seen that 
if there is no ``wave function" (either complex or real) that satisfies 
a linear equation, then the theory of measurement based on 
wave-function collapse is inconsistent.
The possible interpretations of this result are:
\begin{enumerate} 
\item Only CM and linear QM are meaningful physical theories, or
\item More general nonlinear QM is physically meaningful, but a measurement 
cannot be performed, which  is why we cannot measure effects of such 
theories, or
\item The wave-function collapse is not an essential part of measurement 
at all, so 
that all nonlinear generalizations of QM can be interpreted in the same way.
\end{enumerate}
We note that
the only known interpretation that allows consistent interpretation 
for any nonlinear generalization of QM by satisfying 3. is -- 
the Bohmian interpretation.

It is also possible that the conclusions above suffer from our lack of 
imagination. We do not exclude a possibility that someone
may think to have a different (and better) general interpretation of 
nonlinear QM. What we propose is a simple test of the consistency
of any such different interpretation: one should
test it on the classical nonlinear Schr\"odinger equation.
If one does not reproduce the predictions of classical mechanics, 
then this interpretation of nonlinear QM is wrong!

%\bibliographystyle{aipproc}

%\begin{theacknowledgments}
\section*{Acknowledgments}
This work was supported by the Ministry of Science and Technology of the
Republic of Croatia.
%\end{theacknowledgments}

%\endinput

\end{document}